\documentclass[journal]{IEEEtran}
\usepackage{array}
\makeatletter
\newcommand{\thickhline}{%
    \noalign {\ifnum 0=`}\fi \hrule height 1pt
    \futurelet \reserved@a \@xhline
}
\newcolumntype{"}{@{\hskip\tabcolsep\vrule width 1pt\hskip\tabcolsep}}
\makeatother

\ifCLASSINFOpdf
\else
\fi
\usepackage[T1]{fontenc}
\usepackage{flushend}
\usepackage{graphicx}

\usepackage{multirow}
\usepackage{algpseudocode}
\usepackage{algorithm}
\usepackage{color}
\usepackage[table]{xcolor}
\newcounter{inlineenum}
\renewcommand{\theinlineenum}{\alph{inlineenum}}

\newcommand{\ApproxSign}{\raise.17ex\hbox{$\scriptstyle\sim$}}
\usepackage{ctable} 
\usepackage{amsmath}
\usepackage{fancyheadings}
\onecolumn
\begin{document}
\bstctlcite{IEEEexample:BSTcontrol}
\title{Mitigating Read-disturbance Errors in STT-RAM Caches by Using Data Compression}
%
%
%
\author{Sparsh~Mittal
\thanks{S. Mittal is with IIT Hyderabad, India.   e-mail: sparsh0mittal@gmail.com. This aricle is a chapter in the Elsevier book ``Nanoelectronics: Devices, Circuits and Systems'' to be published in 2018.  }
}
\maketitle

\begin{abstract}
Due to its high density and close-to-SRAM read latency, spin transfer torque RAM (STT-RAM) is considered one of the most-promising emerging memory technologies  for designing large last level caches (LLCs).  However, in deep sub-micron region, STT-RAM shows read-disturbance error (RDE) whereby a read operation may modify the stored data value and this presents a severe threat to performance and reliability of STT-RAM caches. In this paper, we present a technique, named SHIELD, to mitigate RDE   in STT-RAM LLCs. SHIELD uses data compression to reduce number of read operations from STT-RAM blocks to avoid RDE and also to reduce the number of bits written to cache during both write and restore operations. Experimental results have shown that SHIELD provides significant improvement in performance and energy efficiency. SHIELD consumes smaller energy than   two previous  RDE-mitigation techniques, namely  high-current restore required read (HCRR, also called restore-after-read) and low-current long latency read (LCLL) and \textit{even an ideal RDE-free STT-RAM cache}. 
 
\end{abstract}


\IEEEpeerreviewmaketitle

\section{Introduction}
Recent trends of increasing core-counts and LLC capacity have motivated researchers to explore low-leakage alternatives of SRAM for designing large LLCs. Due to their high density and near-zero leakage power consumption, non-volatile memories such as  STT-RAM and ReRAM (resistive RAM) have received significant attention in recent years \cite{vetter2015opportunities,kaushik2017next}. Of these,  STT-RAM is considered especially suitable for designing LLCs due to its close-to-SRAM read latency and high write endurance \cite{chun2013scaling}.

Since the write latency/energy of STT-RAM are higher than those of SRAM, previous work has mainly focused on addressing this overhead to enable use of STT-RAM for designing on-chip caches \cite{mittal2014TPDSNVM}. However, a more severe issue of `read disturbance error' (RDE) in STT-RAM caches has not been adequately addressed. Specifically, with feature size scaling, the write current reduces, however, read current does not reduce as much \cite{takemura2010highly}. In fact, for sub-32nm feature size, the magnitude of read current becomes so close to the write current that a read operation is likely to modify the stored data and this is referred to as RDE (refer Section \ref{sec:background} for more details). Since reads happen on critical access path, RDE is likely to severely affect performance and reliability. In fact, it is expected that with ongoing process scaling, \emph{readability} and \emph{not writability} will become the most crucial bottleneck for STT-RAM \cite{mittal2017surveySoftErrorNVM,kang2014readability,zhang2012prospect,ran2015read}.

\textbf{Contributions:} In this paper, we present a technique, named `SHIELD', which shields STT-RAM caches against both RDE and high write latency/energy overhead. SHIELD works by using data compression to reduce the number of bits written to cache (Section \ref{sec:methodology}). Firstly, SHIELD does not write any bits to STT-RAM block for all-zero data, since during next read operation, such data can be reconstructed from compression encoding bits. This avoids RDE and high write latency/energy issue for all-zero data. Secondly, SHIELD keeps two copies of data in the block if the data have compressed width ($CW$) of at most 32B (assuming 64B block size). On the next read operation, one copy gets RDE which is not corrected since the second copy still remains free of RDE. This avoids one restore operation for such narrow data ($0<CW\le32B$). On any future read to this block, the single error-free copy of data is read and then restored. Thirdly, for any data with $CW> 32B$, including an uncompressed data-item, a single copy is written which is restored after each read operation.

\textbf{Salient Features: } By virtue of  writing compressed data, SHIELD reduces the bits written in restore operation and saves write energy. Thus,  SHIELD addresses both RDE and write overhead issue for compressed data, whereas previous techniques \cite{wang2015selective,jiang2016improving} reduce some restore operations only and do not reduce the overhead of read and write operations (Section \ref{sec:features}). SHIELD forgoes the capacity advantage of compression and thus, stores only one cache line in a  block. Hence, SHIELD does not suffer from overheads which generally accompany   compression techniques  \cite{mittal2015compressionSurvey} such as extra tags, compaction, fragmentation, changes in cache replacement policy, etc. Further, some techniques for reducing soft-error vulnerability in SRAM caches  (e.g., \cite{chakraborty2010mc2}) work by keeping two or three copies of an entire cache block (64B) in a cache set. This, however, leads to sharp degradation in cache capacity. By comparison, SHIELD duplicates compressed data of at most 32B size within a 64B cache block itself, and thus SHIELD does not degrade cache capacity. While SHIELD can work with any compression technique, in this paper, we demonstrate SHIELD with base-delta-immediate (BDI) compression algorithm \cite{pekhimenko2012base}.

\textbf{Evaluation and results:} We have performed microarchitectural simulation with an X86 full-system simulator (Section \ref{sec:experimentation}). Also, we have compared SHIELD with two mechanisms for addressing RDE, namely restore-after-read (also called high-current restore required read or HCRR) and low-current long latency read (LCLL) (refer Section \ref{sec:comparisonwithothers}). The results show that SHIELD provides higher performance and energy efficiency compared to these techniques (Section \ref{sec:results}). Compared to an ideal RDE-free STT-RAM LLC, for single and dual-core configurations, SHIELD provides \emph{energy saving} of 1.7\% and 3.0\%, respectively, whereas LCLL (which, on average, performs better than HCRR) incurs \emph{energy loss} of  5.1\% and 5.2\%, respectively. Also, the relative performance with SHIELD is 0.98$\times$ and 0.97$\times$, whereas that with LCLL is 0.96$\times$ and 0.96$\times$, respectively. 
Additional experiments show that SHIELD works well for different cache sizes and selective replication used in SHIELD contributes positively to its effectiveness.

\section{Background}\label{sec:background}
We now provide a brief background and  refer the reader to previous work for more details \cite{takemura2010highly,fong2014failure,mittal2015nvmflashsurvey}. 
\subsection{Motivation for using non-volatile memories}
Conventionally SRAM has been used for designing on-chip caches. However, SRAM has low density and high leakage power due to which large SRAM caches contribute greatly to the overall power consumption. Although several architectural techniques have been proposed to manage the power consumption of SRAM caches, such as cache reconfiguration, near-threshold computing, etc., the power budget targets of next-generation systems require magnitude order higher energy efficiency. This has motivated the researchers to explore alternatives of SRAM, such as eDRAM, STT-RAM, etc. \cite{agarwal2017towards}. 

\subsection{Working of STT-RAM}

STT-RAM  utilizes a Magnetic Tunnel Junction
(MTJ) as the memory storage. An MTJ contains two
ferromagnetic layers separated by an oxide barrier
layer. The magnetization direction of one ferromagnetic layer is fixed while that of the other ferromagnetic layer can be altered by passing a current. The
resistance of the MTJ is determined by the relative
magnetization direction of these two layers. If the two
layers have different directions, the resistance of the
MTJ is high and vice versa. Using this property, a
binary value is stored in an STT-RAM cell \cite{mittal2014TPDSNVM}. 
Another characteristic of STT-RAM is that the read and write operations to it are asymmetric, since a  write operation consumes larger time and energy than a read operation.

Although STT-RAM has lower density than PCM
and RRAM and higher write latency and energy
than SRAM, it has been widely used for designing
caches due to its high write endurance.  Another
advantage of STT-RAM is that its non-volatility can
be traded to improve its write energy and latency. Based on the application
characteristic and the level of cache hierarchy, a designer
can choose a suitable value of retention period.

\subsection{Origin of read-disturbance error}

STT-RAM stores data in magnetic tunnel junction (MTJ) \cite{chun2013scaling}. For performing both read and write operations to MTJ, a voltage is applied between bit line and source line. The only difference between read and write operations is that read operations use smaller voltage than write operations \cite{wang2015selective}. With decreasing feature size, there is an exponential reduction in MTJ area \cite{wang2015selective} such that halving the feature size leads to 25\%  MTJ area and 75\% write current reduction. However, read current does not scale well with feature size scaling since sensing correct data using low-current ($<20\mu A$) is challenging \cite{takemura2010highly,jiang2016improving}. At large feature size (e.g. 130nm), read current is much lower than the write current,  however, at small feature size (e.g. 32nm), read and write current magnitudes become so close that a read may inadvertently write (i.e., disturb) the cell being read. This is referred to as RDE. 

\subsection{Characteristics of read-disturbance error}
RDE is data-dependent, e.g., using read current from BL to SL disturbs cells storing `1' only and not `0'. In STT-RAM, reads and writes cannot be independently optimized, e.g., reducing the MTJ thermal stability for improving switching performance increases probability of read disturbance and thus, RDE mitigation involves tradeoffs. For these reasons, RDE is expected to become the most severe bottleneck in STT-RAM scaling and performance.

It is interesting to note the difference between the  read disturbance error in STT-RAM  and the write-disturbance error (WDE) in PCM \cite{mittal2017surveySoftErrorNVM,wang2017decongest}.  While RDE affects the same cell which is read, WDE affects the nearby cells. Also, RDE happens on a read operation whereas WDE happens on a write operation.

\subsection{Strategies for addressing RDE }
There are several possible strategies for addressing RDE: 
\begin{enumerate}
\item One possible approach for reducing RDEs is to increase STT-RAM write current, which increases the margin between read and write currents. However, since STT-RAM write current and energy values are already high, increasing them further will make STT-RAM unsuitable for use in designing on-chip caches. 
\item 
Conventionally, error-correcting codes are used for mitigating errors in caches \cite{mittal2015reliabilitysurvey,mittal2016reliabilitytradeoffs} and along similar lines, ECC can be used for mitigating read-disturbance error in STT-RAM \cite{seyedzadeh2016leveraging}. However, at small feature sizes, RDE causes very high error rates, such that even with strong error-correcting codes (e.g. 5EC6ED which can correct 5 errors and detect 6 errors), the error rate remains higher than that acceptable for on-chip caches \cite{wang2015selective}.  
 
\item Once the data-value is sensed by sense amplifiers (SAs), it remains locked in SAs and thus, it remains correct (free of RDE) \cite{jiang2016improving}. Hence, after a read operation, writing the stored data back to cells (called `restore' operation) can avoid RDE. This scheme, known as restore-after-read (or HCRR) has been used in a recent STT-RAM prototype \cite{takemura2010highly}. However, this scheme fails to leverage the properties of data value and cache access behavior and hence, incurs large energy and performance penalty (refer Section \ref{sec:mainresults}). 
\item To address RDE,  low-current  sense amplifiers can be used which have larger sensing period. This approach does not create RDE but may increase sensing latency by three times \cite{kang2013high,jiang2016improving}, which harms performance. 
\item Other researchers have proposed architectural management techniques, such as phased access of MRU and non-MRU ways \cite{kong2016novel}, avoiding   restore operations by using LCLL reads \cite{jiang2016improving}, data compression \cite{mittal2017addressing}, data-duplication \cite{mittal2017addressing}, by exploiting cache/register-file access properties \cite{wang2015selective,zhang2016red} and by using an SRAM buffer to absorb reads for minimizing reads from STT-RAM \cite{zhang2016red}. By comparison, some techniques postpone restores by scheduling them at idle times. 

\item  A few other device-level techniques have been proposed for mitigating RDEs \cite{fong2014failure,kang2014variation,ran2015read,bishnoi2014read,raychowdhury2013pulsed,na2016read}, however, these techniques impose  latency, area and/or power overheads \cite{bishnoi2014read} and  their integration into product-system may require addressing several unforeseen challenges. 

 \end{enumerate}

Clearly, architecture-level techniques for mitigating RDEs such as SHIELD can complement device-level techniques and can be especially useful at small feature sizes.

\subsection{Cache properties}
A cache has a set-associative structure, where a newly-arrived cache block  may reside in any of the cache ways, as decided by the cache replacement policy. The optimization target for designing each level of cache is different. The first-level cache (L1) is accessed very frequently and hence, it should have high speed and write endurance, even if it has a small size or high leakage power. The last level cache (L2 in this paper) is designed to reduce off-chip accesses and hence, it must have large capacity. A high latency of last level cache can be usually tolerated using techniques such as instruction-level parallelism. In general, based on their latency and write endurance values, along with capacity advantage, STT-RAM is more suitable for the last level cache than the first level cache. Further, based on the latency-tolerance property of L2 cache, we use data compression in L2 cache and not in L1 cache.

\section{SHIELD: Key Idea and Architecture}\label{sec:methodology}
Before describing the main idea and working of SHIELD in detail, we briefly discuss the compression algorithm used. We also define a metric which is helpful in understanding the effectiveness of any RDE-mitigation technique. 

\subsection{Compression algorithm}\label{sec:compressionalgo}
SHIELD uses data compression approach for reducing the amount of data written to cache. While SHIELD can work with any compression technique, in this paper, we illustrate its working using BDI (base-delta-immediate) compression algorithm \cite{pekhimenko2012base} due to its low decompression latency and reasonable compression ratio. 

It is well-known that for many cache blocks, the difference between the values stored in the blocks is small \cite{mittal2015compressionSurvey}. Based on this, BDI represents a block with a base and an array of differences, which reduces the total block size compared to the original size. To compress blocks with two different dynamic ranges, BDI uses two bases where one base is always zero and the second base is taken from actual data-contents \cite{pekhimenko2012base}.  BDI algorithm first tries to compress every element using a zero base. Then, for any uncompressed block, compression is attempted by taking the first uncompressed element as the base.   A 1-bit mask is used to record whether the corresponding base is zero.  

A BDI-compressed block is represented as  $\text{B}_p\Delta _q$, where  $p$ and $q$ represent the size of base and $Delta$ (difference), respectively. BDI uses 6 such patterns, viz. $\text{B}_8\Delta _1$, $\text{B}_8\Delta _2$, $\text{B}_8\Delta _4$, $\text{B}_4\Delta _1$, $\text{B}_4\Delta _2$ and $\text{B}_2\Delta _1$. BDI also checks if the block has repeated 8-byte values or is all-zero \cite{pekhimenko2012base}. Thus, after attempting to compress the block data using these 8 compression states, the one providing the smallest compressed size is finally chosen.

\subsection{Defining consecutive reads}\label{sec:creadillustration}
\begin{figure}[ht]
 \centering
 \includegraphics [scale=0.44] {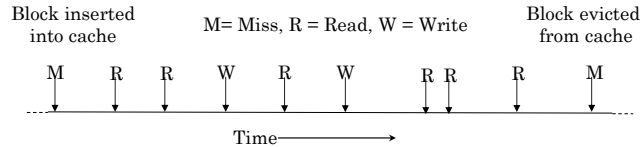} 
 \caption{One generation of a cache block (for defining CRead).} 
  
\label{fig:creadillustration}
 \end{figure}

For any cache block, we define a metric \emph{consecutive-read (CRead)}, which measures the average number of consecutive reads (i.e., no intermediate writes) seen by it during its residency in LLC. Also, for any application, the  CRead is defined as the average CRead for all its cache blocks.  Note that CRead is defined differently than the `reuse-count' of a block which measures both read and write operations in one generation of a block. For example, in Figure \ref{fig:creadillustration}, CRead for the block is 2 (= (2+1+3)/3), whereas reuse-count is 8. 

To see the significance of CRead, assume a hypothetical application where every LLC read is followed by a write operation and thus, CRead=1. Since an actual write operation automatically corrects any RDE-affected data, an oracle scheme can completely avoid restore operations for this application and thus, this application does not need restore operations. As another example, if every write is followed by 10 reads on average, then CRead=10. In this case,  automatic correction through write operations happens infrequently and larger number of restore operations are required. Clearly,  higher the value of CRead, higher is the restore requirement of an application and vice versa.

\subsection{SHIELD: Key Idea}\label{sec:keyidea}
Caches work on the temporal locality principle, which states that a data item which is referenced at one point of time will be referenced again in near future. Such data item is stored in caches to avoid accessing main memory.  For an RDE-affected STT-RAM cache, data are written to cache on both actual write operations and restore operations.  To reduce the data written to STT-RAM during both these  operations, SHIELD  writes data in compressed form using BDI algorithm. This reduces the STT-RAM cache write energy. We  perform two optimizations to BDI algorithm \cite{mittal2016softErrorGLSVLSI,mittal2017addressing} for avoiding restore operations and further reducing the amount of data written to cache. 
 
First and importantly, for data with all-zeros, no bit is stored in the STT-RAM block since the original uncompressed data can be easily recovered based on compression state encoding bits. Second, the first delta in BDI algorithm is always zero since it is the difference between the non-zero base with itself. Hence, we do not store this delta. The corresponding element is generated from the non-zero base. 

Based on our optimizations and LLC access pattern, SHIELD uses the following insights to reduce restore operations.

\textbf{1.} For all-zero data ($CW=0$), no bits are written to or read from STT-RAM blocks and thus, RDE does not occur for such data. Hence,  \emph{all} the restore operations to such data are completely avoided.

\textbf{2.} It is well-known that due to filtering from L1 cache, LLC sees much smaller data locality than L1 cache \cite{mittal2014surveycache}. Hence, most LLC blocks are read only few times. In fact, our experiments have shown that average value of CRead for single and dual-core configurations is 1.61 and 1.32, respectively (refer Figures \ref{fig:results1core}(c) and \ref{fig:results2core}(c)). Clearly, after any write operation, cache blocks see less than 2 read operations. Based on this, SHIELD keeps two copies of the data with $0<CW\le32B$ within the same block.  On a later read operation, one copy gets RDE, whereas the second copy still remains free from RDE. Thus, keeping two copies is equivalent to restoring the data at the time of original write operation to avoid \emph{one} future restore operation. Thus, if CRead value is less than two, keeping two copies can avoid at least half of the restore operations for such narrow-width ($0<CW\le32B$) blocks.

 \begin{figure}[ht]
 \centering
 \includegraphics [scale=0.46] {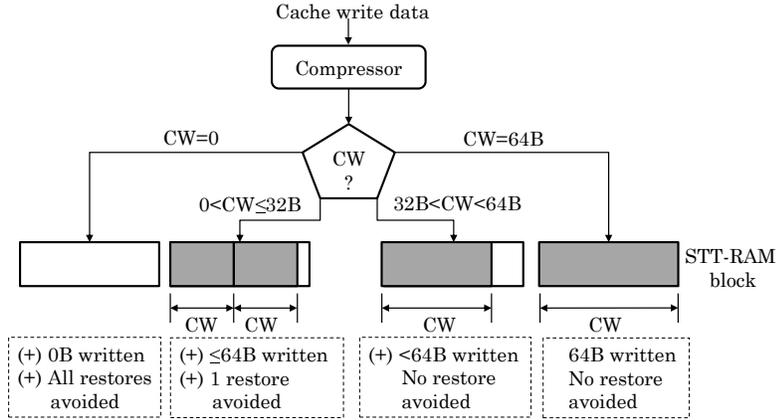}
   \caption{Overview of SHIELD and action taken on a cache write operation}
\label{fig:SHIELDoverview}
 \end{figure}

\subsection{Action on read and write operations}
We now discuss the action taken by SHIELD on any write and read operation.

\textbf{Write operations:} On any write operation, the cache controller compresses the data and computes CW. For $0<CW\le32$, two copies of data are stored and for other $CW$ values, only one data copy is stored. Figure \ref{fig:SHIELDoverview} summarizes the action taken on a write operation and benefit/cost incurred for different $CW$ values.
Table \ref{tab:compressedsize} shows the 4-bit encoding used for different compression states for BDI algorithm. Out of 16 combinations from 4 bits, 13 combinations are utilized and the remaining 3 are not utilized. For each block, this 4-bit encoding  is  stored at the time of a write operation using a memory technology that does not suffer from RDE (e.g. SRAM). Note that due to our optimizations to BDI algorithm (refer Section \ref{sec:keyidea}), the sizes of compressed blocks are different in our work (shown in Table \ref{tab:compressedsize}) than those in the original BDI algorithm \cite{pekhimenko2012base}. 

\begin{table}[htbp] 
\centering
\footnotesize
\caption{Encoding and compressed sizes for different BDI compression states, all sizes are in bytes (Enc. = encoding, Uncomp. = uncompressed) }
 \setlength\tabcolsep{3pt}%
\begin{tabular}{|c|c|c|c"c|c|c|c| }\hline
Enc.  &  State   &     Copies & Size  &  Enc.  &  State   &  Copies   & Size      \\\hline
    0000  &  Zeros  & 1     & 0     & 1100  &  $\text{B}_4\Delta _1$ & 1 & 19     \\\hline
    0001  &  Repeat  & 1     & 8     & 1101  &  $\text{B}_4\Delta _1$ & 2 & 19$\times$2   \\\hline
    0011  &  Repeat  & 2 & 8$\times$2        & 0100  &  $\text{B}_4\Delta _2$ & 1    & 34 \\\hline
    0010  &  $\text{B}_8\Delta _1$ & 1 & 15         & 1110  &  $\text{B}_2\Delta _1$ & 1 & 33    \\\hline
    0110  &  $\text{B}_8\Delta _1$ & 2 & 15$\times$2       & 1000  &  $\text{B}_8\Delta _4$ & 1 & 36  \\\hline
    0101  &  $\text{B}_8\Delta _2$ & 1 & 22       & 1111  &  Uncomp. & 1 & 64   \\\hline
    0111  &  $\text{B}_8\Delta _2$ & 2 & 22$\times$2        &       &       &       &  \\\hline
\end{tabular}%
\label{tab:compressedsize}%
\end{table}%
   
\textbf{Read operations:} On any read operation, first the encoding (E) is consulted. If  $E=0000$, then the zero-data are  reconstructed without accessing the block which avoids  RDE and need of restoration. If $E$ $=$ $0011$, $0110$, $1101$ or $0111$, it implies that $0<CW\le32B$ and 2 copies of data are stored. Hence, one copy is read,  no restore is performed and the encoding is changed to $0001$, $0010$, $1100$ and $0101$, respectively, which indicates that now only one (error-free) copy of data is stored in the block. If $E$ $=$ $0001$, $0010$, $1100$, $0101$, $0100$, $1110$, $1000$ or $1111$, then data value is read, a restore operation is issued and the encoding remains unchanged.



\subsection{Overhead assessment}\label{sec:overhead}
\textbf{Latency overhead:}  The latency overhead of BDI compression and decompression is 2 and 1 cycles, respectively \cite{pekhimenko2012base}. We account for these values in performance simulation. Since STT-RAM write latency is high, a small value of BDI compression latency is easily hidden. 
  
\textbf{Energy overhead:} We now compute the energy overhead of compression and decompression. From Wu et al. \cite{wu2006analysis}, we note that a 1-byte (8-bit) adder consumes nearly 15 fJ energy. Assuming that a subtractor also consumes same amount of energy, we first compute the overhead of compression by counting the number of subtractions in BDI algorithm. Each of $\text{B}_8\Delta _1$, $\text{B}_8\Delta _2$ and $\text{B}_8\Delta _4$ perform seven 8-byte subtractions. Both $\text{B}_4\Delta _1$ and $\text{B}_4\Delta _2$ perform fifteen 4-byte subtractions and $\text{B}_2\Delta _1$ performs thirty-one 2-byte subtractions.  Repeated value detection requires seven 8-byte subtractions.  Thus, there are 406 (=3$\times$7$\times$8+2$\times$15$\times$4+1$\times$31$\times$2+7$\times$8) byte subtractions, which consume 6.09 pJ (=406$\times$15 fJ). To account for other overheads and zero detection circuit, we increase the value and assume the overhead of compression as 8 pJ. As for decompression, it requires a maximum of thirty-one 2-byte additions, which consume 0.93 pJ (=31$\times$2$\times$15 fJ). Hence, we assume the decompression overhead as 1 pJ. We account for these overheads in energy computations. 

To see the energy overhead of encoding bits, we note that a 2-bit counter consumes 0.1pJ in each access \cite{mittal2015equalwrites}, thus a 4-bit counter would consume 0.20pJ. Since the encoding is written at the time of a cache write, we compare this energy with the write energy of STT-RAM cache. From Table \ref{tab:l2parameters}, write energy of 4MB L2 cache is 0.389nJ and thus, the dynamic energy of encoding bits is 0.05\% of the cache energy. Hence, we ignore its contribution in energy.

\section{Salient Features of SHIELD and Qualitative Comparison}\label{sec:features}

Some researchers have proposed techniques for addressing RDEs in STT-RAM caches. 
The technique of Wang et al. \cite{wang2015selective} seeks to avoid restore operations for blocks which are expected to see a write operation in near future. To perform L2 restore operations and updating L1 or L2 metadata bits, their technique requires observing the state of the block in other (i.e., L2 or L1) cache. Also, their technique postpones restore operations and hence, in their cache design, a dirty L2 block may have RDE. Since this RDE-affected L2 block cannot be written back to memory, their technique writes back the corresponding block in L1 cache directly to memory. Due to these, their technique complicates cache management and incurs overhead which increases with rising number of cores.  Also, their technique would require significant modifications to work with different cache coherence schemes, however, they do not present these details. By comparison, SHIELD performs restore operations immediately and hence, always keeps the L2 cache RDE-free.  Also, SHIELD works based on data-width only and hence, can be integrated with any cache coherence scheme.

Jiang et al. \cite{jiang2016improving} propose selectively performing short-latency read (requires restore) and LCLL read  (long-latency read with no requirement of restores) depending on whether a particular bank is idle. However, short-latency reads still require restores and LCLL reads increase read latency. A key benefit of SHIELD is that it avoids read, write and restore operations for all-zero data and restore operations for narrow data. Thus,  SHIELD addresses both write latency/energy overhead and RDE issue by using data compression. By comparison, previous techniques \cite{wang2015selective,jiang2016improving} reduce some restore operations only and do not reduce read operations or address the overhead of write operations. Also, the technique of Jiang et al. \cite{jiang2016improving} is proposed for main memory where the memory controller buffers read/write requests. Due to differences in cache and main memory architecture, their technique may not be applicable or effective for caches.   

To reduce the restore energy, some researchers propose restoring only cells with `1' value \cite{wang2015selective}, since only these cells may be disturbed during reads. These techniques are orthogonal to SHIELD and hence, can be easily integrated with it. 

Note that SHIELD does not use an additional buffer for postponing refresh operations and performs any required restore operation immediately after a read operation. Thus, the data in the cache are always kept error-free. Also, SHIELD actually reduces restore operations, whereas techniques that use a buffer for postponing restore operations do not reduce restore operations, but merely reschedule them to avoid interfering with normal accesses.

\section{Experimentation Platform}\label{sec:experimentation}
\subsection{Simulator parameters}
We use Gem5 simulator to perform simulations using detailed timing model. L1 data/instruction caches have  32KB size with 2-way associativity.  The L1 caches are private to core and the L2 cache (LLC) is shared between cores. The size of L2 cache in single and dual-core configurations is 4MB and 8MB, respectively. All caches are write-back and use LRU replacement policy. The L2 cache parameters for 16-way STT-RAM L2 are obtained using DESTINY \cite{poremba2015destiny,mittal2014destiny,mittal2017DESTINY} for 32nm feature size.  We assume that the cache is optimized for write EDP (energy-delay-product) and uses sequential tag-data access. The values obtained are shown in Table \ref{tab:l2parameters}.  
  
\begin{table}[htbp]
\footnotesize
  \centering
  \caption{STT-RAM L2 cache parameters}
    \begin{tabular}{|c|c|c|c|c|}\hline
    & 2 MB & 4 MB & 8 MB & 16 MB \\\hline
    Hit latency (ns) & 4.063 & 3.737 & 4.058 & 4.350 \\
    Miss latency (ns) & 1.976 & 1.567 & 1.805 & 1.814 \\
    Write latency (ns) & 4.920 & 4.970 & 5.003 & 5.145 \\\hline
    Hit energy (nJ)  & 0.264 & 0.304 & 0.333 & 0.391 \\
    Miss energy (nJ) & 0.107 & 0.105 & 0.112 & 0.113 \\
    Write energy (nJ) & 0.366 & 0.389 & 0.427 & 0.490 \\
    Leakage power (W) & 0.019 & 0.044 & 0.072 & 0.138 \\\hline
    \end{tabular}%
  \label{tab:l2parameters}%
\end{table}%

\subsection{Workloads}

We use all the 29 benchmarks from SPEC2006 suite with \textit{reference} inputs and 3 benchmarks from HPC field  (shown as italics in Table  \ref{tab:workloads}) as single-core workloads. Using these, we randomly create 16 dual-core multiprogrammed workloads, such that  a benchmark is used exactly once. Table \ref{tab:workloads} shows the workloads.

\begin{table}[htbp]
\footnotesize
\centering
\caption{Workloads used in the paper} 
\begin{tabular}{|c|c|}
\hline 
 Single-core workloads and their acronyms \\\hline
 As(astar), Bw(bwaves), Bz(bzip2), Cd(cactusADM)  \\
  Ca(calculix), Dl(dealII), Ga(gamess), Gc(gcc) \\
   Gm(gemsFDTD), Gk(gobmk), Gr(gromacs), H2(h264ref)\\
    Hm(hmmer), Lb(lbm), Ls(leslie3d), Lq(libquantum)\\
    Mc(mcf), Mi(milc), Nd(namd), Om(omnetpp) \\
   Pe(perlbench), Po(povray), Sj(sjeng), So(soplex)\\
   Sp(sphinx), To(tonto), Wr(wrf), Xa(xalancbmk)  \\
   Ze(zeusmp), \textit{  Co(comd)}, \textit{Lu(lulesh)},   \textit{Xb(xsbench)} \\\hline
 Dual-core workloads (using  acronyms shown above) \\\hline
 BwLu, PeLq, XaTo, CaMc, GkLb, GmWr, PoSp, NdGr,\\
 GcMi, BzZe, LsSo, SjXb, OmH2, CoCd, AsDl, HmGa \\\hline
\end{tabular}
\label{tab:workloads}
\end{table}
 
\subsection{Simulation completion strategy } 
In multiprogrammed workloads, different programs have different IPCs and thus, different rate of progress. Hence, they may execute the same number of instructions in different amount of time. Due to this, a strategy is required for ensuring that in each run with a workload (with baseline or a technique), the number of instructions simulated should be nearly equal.

One strategy for this is to simulate each workload for a fixed number of cycles, however, in this strategy, the number of instructions simulated may be different for baseline and a technique, depending on the performance impact of a technique. 

We use another strategy, where we simulate every workload till each application in the workload has  executed nInst instructions. The value of nInst is 150M for single-core workloads and 100M for dual-core workloads. This keeps the simulation turnaround time manageable since we perform detailed timing simulation with a full-system simulator and simulate a large number of workloads, techniques (baseline, SHIELD, HCRR, LCLL) and their parameter configurations/variants. The early-completing benchmarks continue to run but their IPC (instruction per cycle) is recorded only for the first nInst instructions, following the well-established simulation methodology \cite{mittal2016technique,mittal2016softErrorGLSVLSI}. Remaining metrics are computed for the whole execution, since they are system-wide metrics (whereas IPC is a per-core metric). 

\subsection{Comparison with related schemes}\label{sec:comparisonwithothers}
Our baseline is an ideal scheme that assumes a cache free of  RDE. We compare our technique with two other schemes for mitigating RDE.

\textbf{High current restore required read (HCRR) :} This technique uses normal current ($20\mu A$) to read the STT-RAM cell and performs a restore (i.e. write)  operation after every read operation to correct the RDE. This technique is also referred to as refresh (or restore) after read \cite{wang2015selective} and `disruptive reading and restoring' \cite{bishnoi2014read} technique. 
 
\textbf{Low-current long latency read (LCLL):} To avoid RDEs, low-current (\ApproxSign $10\mu A$) sense amplifiers (SAs) can be used which have larger sensing duration \cite{kang2013high,jiang2016improving}. These SAs also require extra sensing stages to minimize sensing margin degradation due to variations in STT-RAM cells. This approach, referred to as LCLL, does not create RDE, but incurs 3$\times$ the sensing latency of traditional SAs \cite{kang2013high,jiang2016improving}. 
 \begin{figure*}[ht]
 \centering
 \includegraphics [scale=0.44] {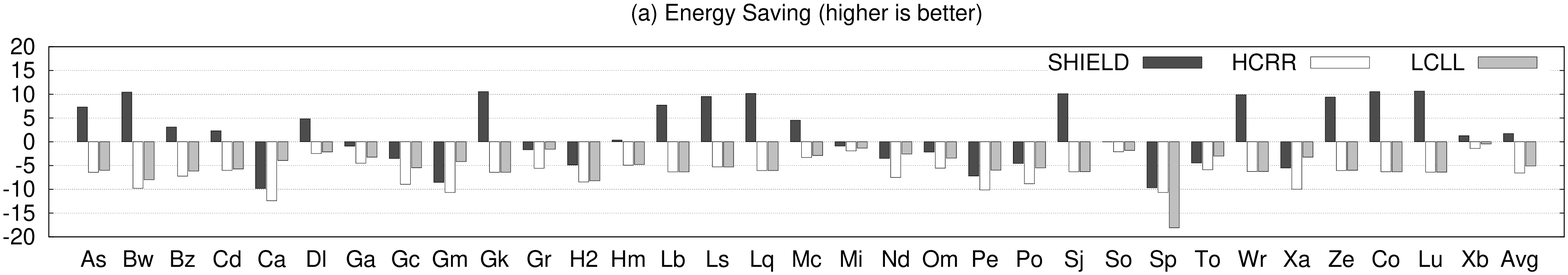}
 \includegraphics [scale=0.44] {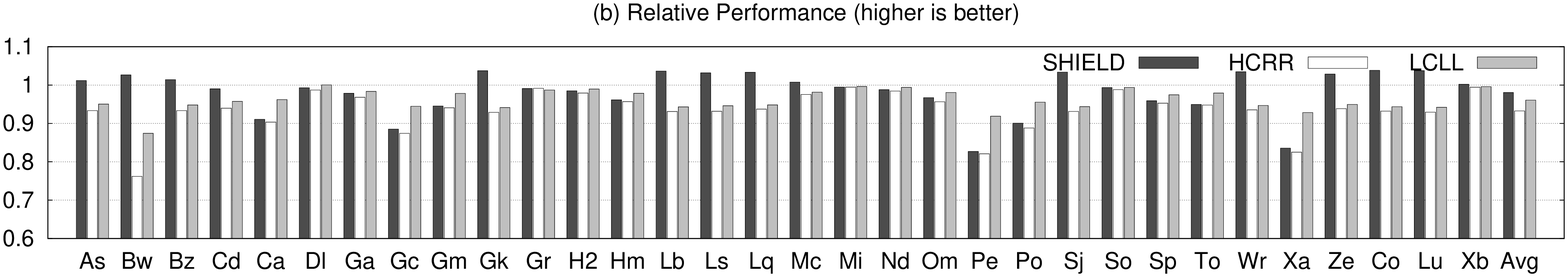}
 \includegraphics [scale=0.44] {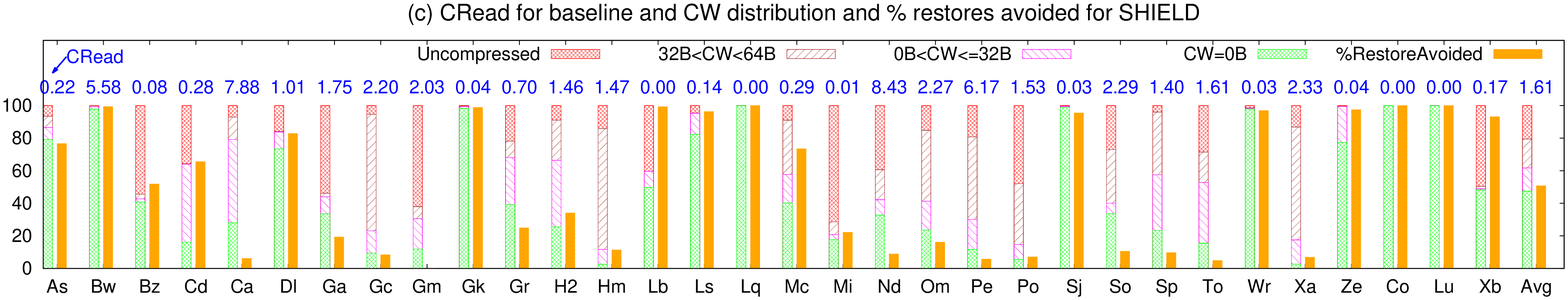} 
 \includegraphics [scale=0.44] {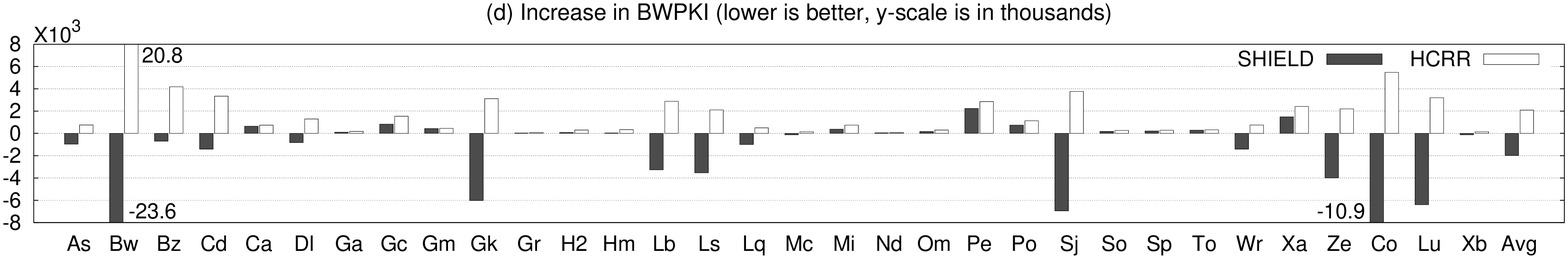} 
   \caption{Results for single-core system}
\label{fig:results1core}
 \end{figure*} 

\subsection{Evaluation Metrics}
\begin{figure*}[ht]
 \centering
 \includegraphics [scale=0.44] {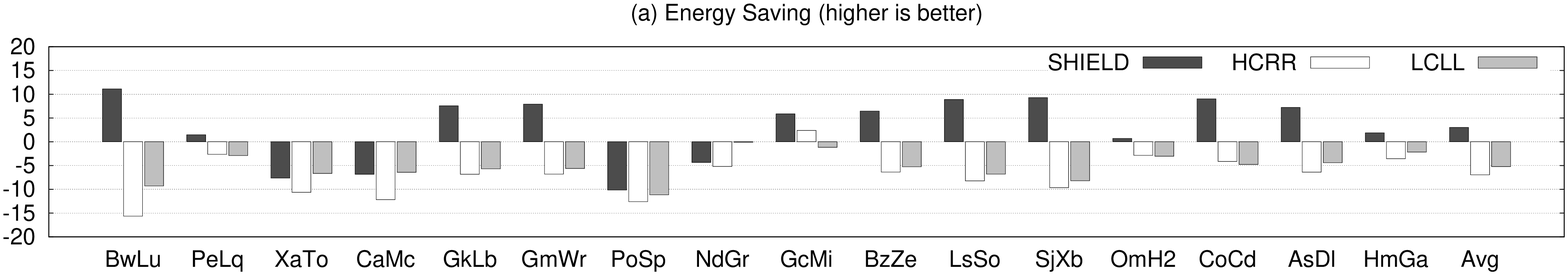}
 \includegraphics [scale=0.44] {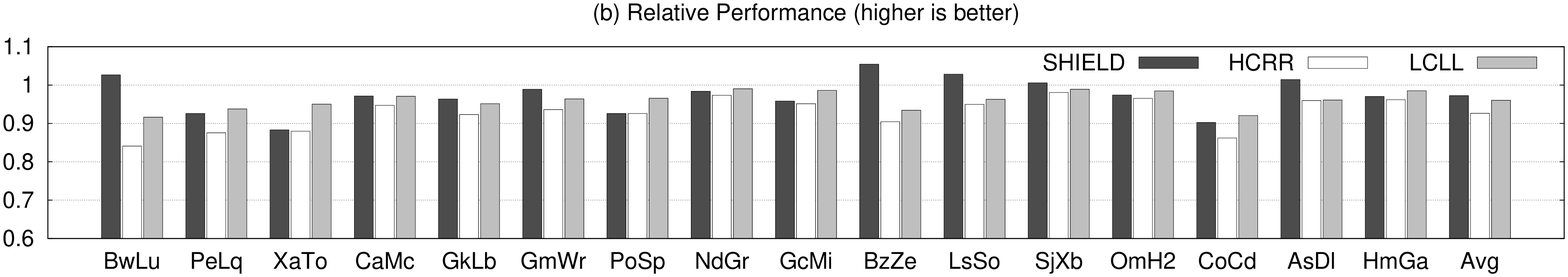}
 \includegraphics [scale=0.44] {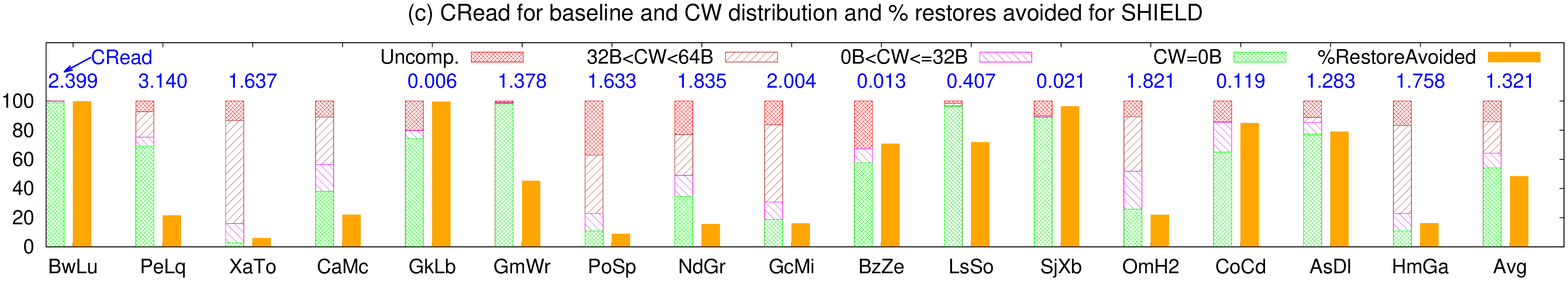} 
 \includegraphics [scale=0.44] {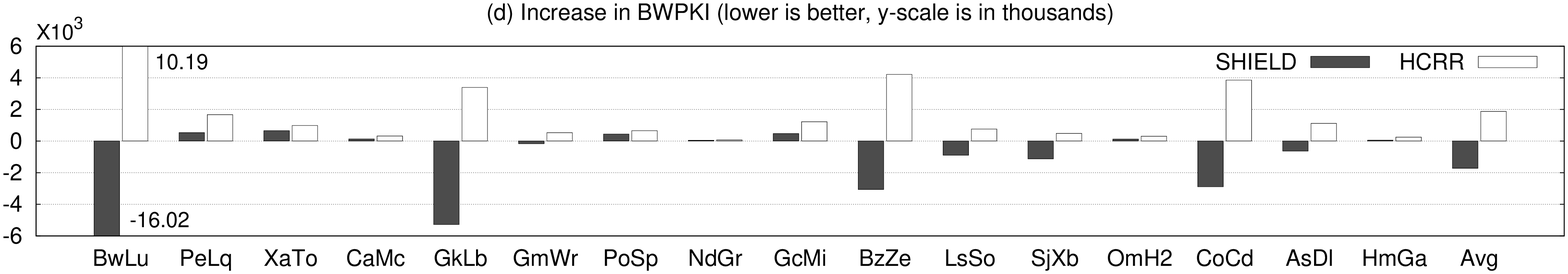}
   \caption{Results for dual-core system}
\label{fig:results2core}
 \end{figure*}

We use the following evaluation metrics:
\begin{enumerate}
\item  L2 cache energy (leakage+dynamic)
\item  Weighted speedup (referred to as relative performance), defined as $\Sigma_{n}
(\text{IPC}_n(\text{technique})/\text{IPC}_n(\text{baseline}))/N$, where $N$= number of cores.  
\item  Average CRead value  for baseline (Section  \ref{sec:creadillustration}). 
\item $\Delta$BWPKI (=BWPKI$_{\text{technique}}$-BWPKI$_{\text{baseline}}$), where BWPKI refers to bytes written to cache per kilo instruction. BWPKI shows the write traffic to cache and since compression changes the size of each block, we have used BWPKI instead of writes per kilo instruction. Thus, $\Delta$BWPKI shows the increase in write traffic due to RDE under any technique (SHIELD or HCRR) compared to an ideal RDE-free cache. Since this metric for LCLL is nearly zero, we do not show this value for LCLL in Figures \ref{fig:results1core}(d) and \ref{fig:results2core}(d). 
\end{enumerate}

For SHIELD, we also show  
\begin{enumerate}
\setcounter{enumi}{4}
\item \emph{Compressed width ($CW$) of data on each cache write}. We classify the $CW$ values in four ranges (refer Table \ref{tab:compressedsize}): $CW=0$,  $0<CW\le32B$ (i.e. \{8B,15B,19B,22B\}), $32B<CW<64B$ (i.e. \{33B,34B,36B\}), $CW=64B$ (i.e. uncompressed).
\item Percentage of restore operations avoided (RstAvd). Let $Reads_{CW=0}$ be the reads to blocks with $CW=0$ data and $Reads_{0< CW \le32\_2copy}$ be the reads to blocks with $0<CW\le32B$ data having two copies. Then,  $RstAvd= \dfrac{(Reads_{CW=0} +Reads_{0< CW \le32\_2copy})\times 100}{TotalReads}$ 

Note that for BDI algorithm, $Reads_{CW=0}$ corresponds to $E$ $=$ $0000$ and $Reads_{0< CW \le32\_2copy}$ corresponds to $E$ $=$ $0011$, $0110$, $1101$ or $0111$.

\end{enumerate}

 \section{Results and Analysis}\label{sec:results}
\subsection{Main Results}\label{sec:mainresults}
Figure \ref{fig:results1core} and \ref{fig:results2core} shows the results and Table \ref{tab:resultssummary} summarizes the results.  We now analyze the results.

\begin{table}[htbp]
\footnotesize
  \centering
  \caption{Summary of results (rel. perf. = relative performance)}
  \setlength\tabcolsep{3pt}%
    \begin{tabular}{|c|c|c|c"c|c|c|}\hline
          & \multicolumn{3}{c"}{Single-core} & \multicolumn{3}{c|}{Dual-core} \\\hline
          & SHIELD & HCRR  & LCLL  & SHIELD & HCRR  & LCLL \\\hline
    Energy Saving & 1.73\%  & -6.57\% & -5.09\% & 3.03\%  & -6.96\% & -5.23\% \\\hline
    Rel. Perf. & 0.98$\times$  & 0.93$\times$  & 0.96$\times$  & 0.97$\times$  & 0.93$\times$  & 0.96$\times$ \\\hline
    $\Delta$BWPKI & -1980 & 2085  & -4    & -1725 & 1877  & -5 \\\hline
    \end{tabular}%
  \label{tab:resultssummary}%
\end{table}%
  
\textbf{SHIELD:} Compared to the RDE-free cache (baseline), both HCRR and LCLL incur \emph{energy loss}, whereas SHIELD provides  \emph{energy saving}. Also, SHIELD provides higher performance than HCRR and LCLL. The energy  and performance improvement of SHIELD for an application depends on the fraction of restores avoided. This, in turn, depends on  the compressibility of application's data blocks and its read/write access pattern, specifically its CRead value. Out of total 48 workloads, compared to baseline, SHIELD saves energy for 28 workloads and improves performance for 18 workloads.

\emph{Results on percentage of restores avoided:}  For single and dual-core configurations, SHIELD  reduces 50.7\% and 48.5\% of restore operations, respectively. For many applications, more than 95\% of restores are avoided, e.g. Bw(99.3\%), Gk(98.8\%), Lb(99.3\%), Ls(96.3\%), Lq(99.9\%), Sj(95.4\%), Wr(97.0\%), Ze(97.4\%), Co(99.9\%), Lu(99.9\%), BwLu(99.7\%), GkLb(99.4\%), SjXb(96.3\%), etc. This happens because either these applications have highly compressible data and/or their CRead values are small. 
 
\emph{Data compressibility and CRead results:} For some applications, more than 95\% of blocks are all-zero ($CW=0$), e.g. Bw(97.8\%), Gk(98.2\%), Lq(99.9\%), Sj(98.9\%), Wr(97.9\%), Co(99.9\%), Lu(99.9\%), BwLu(98.7\%), GmWr(98.2\%), etc.  Similarly, low CRead values for Gk(0.04), Lb(0.0003), Lq(0.0003), Sj(0.03), Wr(0.02), Ze(0.03), Co(0.002), Lu(0.00006), GkLb(0.006), BzZe(0.01), SjXb(0.02) etc., indicate that only few restore operations are required for these applications since SHIELD avoids restores for narrow-data ($0<CW\le32B$) and these applications are  read-unintensive. For these applications, large performance improvement and energy saving are achieved and write traffic is significantly reduced.
 
For some applications CRead values are small, (e.g.,  Mi(0.006), etc.),  but only few blocks are compressible. Opposite is true for applications where CRead values are high, (e.g. Bw(5.58), BwLu(2.40), PeLq(3.14), etc.), but most blocks are compressible.  
From  perspective of energy saving with SHIELD, these effect partially cancel each other, and hence, these applications show only small energy gain/loss.  

For some applications, CRead values are  high,  e.g. Gc(2.20392), Gm(2.0307), Ca(7.88), Nd(8.42531), Om(2.27), Pe(6.16507), So(2.286), Xa(2.32), etc. The restore requirement for these applications is also high and thus, restore operations lead to performance and energy loss for these applications. On using SHIELD, only four out of 48 workloads show relative performance less than 0.90$\times$ viz. Gc(0.89), Pe(0.83), Xa(0.84) and XaTo (0.88$\times$). Similarly, in some  applications, most blocks are either incompressible or have compressed width greater than 32B, e.g., Gc, Gm, Hm, Mi, Pe, Po, To, Xa,  XaTo, PoSp, HmGa, etc., and hence,  compared to RDE-free baseline, SHIELD incurs energy loss or very small energy saving due to restore operations.

\emph{$\Delta$BWPKI results:} SHIELD brings large reduction in cache write traffic, which is due to the use of compression and avoiding many restore operations. Especially for applications with compressible data, SHIELD brings large  reduction in BWPKI, e.g., Bw(23606), Gk(6011), Sj(6945),  Co(10942), Lu(6397), BwLu(16015), GkLb(5281) etc. Clearly, in addition to RDE, SHIELD also addresses  the write overhead in STT-RAM caches.

\emph{Overhead of (de)compression:} For single and dual-core systems, the compression and decompression energy (refer Section \ref{sec:overhead}) together account for 0.14\% and 0.15\% of the total L2 energy. Clearly, the energy overhead of SHIELD is negligible. 

We conclude that SHIELD can bridge the large performance gap between an ideal RDE-free cache and an RDE-affected cache that uses restore-after-read (HCRR) scheme. Also, SHIELD completely removes the energy overhead of restore operations and hence, it has lower energy consumption and lower LLC write traffic than an ideal RDE-free cache. 

\textbf{HCRR:} With HCRR, each read operation also leads to a write operation and thus, HCRR causes significant increase in write traffic to STT-RAM cache. The increase in BWPKI is especially high for read-intensive applications, e.g. Bw(20844), Bz(4181), Sj(3766), Co(5478), BwLu(10194), BzZe(4218), CoCd(3858), etc. Since STT-RAM writes have high energy/latency overhead, HCRR incurs large energy and performance loss and is also likely to cause bandwidth issues. In fact, the worst case performance with HCRR can be very poor, e.g.   Bw(0.76$\times$), Gc(0.87$\times$), Pe(0.82$\times$), Po(0.89$\times$), Xa(0.83$\times$), BwLu(0.84$\times$), PeLq(0.88$\times$), XaTo(0.88$\times$), CoCd(0.86$\times$), etc.  Also, for some applications, HCRR incurs more than 10\% energy loss, e.g., Ca(-12.4\%), Gm(-10.7\%), Pe(-10.1\%), Sp(-10.6\%), XaTo(-10.6\%), CaMc(-12.1\%), PoSp(-12.6\%), etc. With increasing number of cores, the LLC access intensity increases and hence, restore operations will cause port obstruction and make the LLC unavailable for serving accesses. Clearly, due to its large penalty, use of HCRR in performance-critical systems is challenging.

\begin{table*}[htbp]
  \centering
  \caption{Parameter Sensitivity Results (RstAvd = \% restores avoided)}
    \begin{tabular}{|c|c|c|c|c"c|c|c|c|}\hline
          & \multicolumn{4}{c"}{Single-core}     & \multicolumn{4}{c|}{Dual-core} \\\hline
          & Energy Saving & Rel. Perf.  & $\Delta$BWPKI & RstAvd & Energy Saving & Rel. Perf.  & $\Delta$BWPKI & RstAvd \\\hline
    Default & 1.73\%  & 0.981$\times$ & -1980 & 50.7\%  & 3.03\%  & 0.973$\times$ & -1725 & 48.4\% \\\specialrule{.1em}{.05em}{.05em}  
    SHIELD1 & 1.84\%  & 0.978$\times$ & -2003 & 45.3\%  & 3.16\%  & 0.967$\times$ & -1726 & 43.9\% \\\hline
    SHIELD3 & 0.79\%  & 0.980$\times$ & -1938 & 50.4\%  & 2.70\%  & 0.972$\times$ & -1692 & 48.5\% \\\specialrule{.1em}{.05em}{.05em}  
    Half-size L2  & 2.84\%  & 0.980$\times$ & -1983 & 50.3\%  & 3.77\%  & 0.983$\times$ & -1799 & 48.1\% \\\hline
    Double-size L2 & 1.1\%   & 0.977$\times$ & -1970 & 51.0\%  & 3.39\%  & 0.992$\times$ & -1710 & 49.0\% \\\hline
    \end{tabular}%
  \label{tab:sensitivityresults}%
\end{table*}%

\textbf{LCLL:} LCLL degrades performance by slowing down the read operations. Since read operations happen on critical access path, the large read latency in LCLL may not be easily hidden. In worst case, LCLL can cause large energy loss, Sp(-18.1\%), PoSp(-11.2\%). On average, LCLL incurs energy loss for both single and dual-core configurations, whereas SHIELD saves energy. Compared to RDE-free cache (baseline), LCLL has negligible impact on BWPKI, whereas SHIELD brings large reduction in BWPKI (as shown above), thus, SHIELD  reduces the write traffic to cache more effectively than LCLL. 

Although LCLL provides better performance and energy efficiency compared to HCRR, a crucial limitation of LCLL is that use of small read current in LCLL can lead to \emph{decision failure} \cite{bishnoi2014read}, such that it may not be possible to distinguish between two states of a bit-cell during read operation. Also, since process variation affects device parameters \cite{mittal2016processvariation}, to guarantee RDE-free read operations, the  value of read current needs to be set even lower than that assumed here \cite{zhang2012prospect}; this, however, would further increase the read latency and cause performance loss. Given this, SHIELD presents as a better technique for mitigating RDE than LCLL.

\subsection{Parameter Sensitivity Results}
 We henceforth focus exclusively on SHIELD. Also, we omit per-workload results for brevity and only present average results in Table \ref{tab:sensitivityresults}. For comparison purpose, the results with default parameters are also shown.
\subsubsection{Effect of Replication}
As shown in Section \ref{sec:keyidea}, SHIELD replicates the narrow ($0<CW\le32B$) data to reduce restore operations. We now present two variants of SHIELD which explore the tradeoffs associated with it.

\textbf{SHIELD1:} SHIELD1 does not replicate a data-item with width $0<CW\le32B$, but otherwise works same as SHIELD. For BDI algorithm, four states, viz.,  repeated value, $\text{B}_8\Delta _1$, $\text{B}_8\Delta _2$ and $\text{B}_4\Delta _1$ have $CW$ at most 32B. 

SHIELD1 reduces the number of bits written originally at the cost of incurring one extra restore operation later. Due to this, SHIELD1 saves higher energy  but achieves lower performance than SHIELD (Table \ref{tab:sensitivityresults}) and thus, a designer can choose an appropriate variant to optimize a metric of interest. It is noteworthy that the higher performance of SHIELD can reduce execution time which leads to lower leakage power in other processor components and hence, SHIELD may be preferable over SHIELD1. Also, SHIELD avoids larger percentage of restore operations than SHIELD1. For single and dual-core systems, $0<CW\le32B$ data account for 14.3\% and 10.0\% of data blocks (refer Figures \ref{fig:results1core}(c) and \ref{fig:results2core}(c)) and thus, due to the relatively smaller contribution of $0<CW\le32B$ data, the difference between performance/energy of SHIELD and SHIELD1 is relatively small. 

\textbf{SHIELD3:} On any write operation, SHIELD3 makes three  copies of data with   $0<CW<22B$, but otherwise works same as SHIELD (i.e., two copies of data with $0<CW\le32B$ and one copy for other data). For BDI algorithm, three states, viz., repeated value, $\text{B}_8\Delta _1$, and $\text{B}_4\Delta _1$ have $CW$ at most 22B.

SHIELD3 avoids  the need of restore on two read operations at the cost of increasing the bits written to cache initially. However, SHIELD3 reduces energy saving without providing corresponding performance improvement. This is because average CRead values for single and dual-core configurations is 1.61 and 1.32, respectively (refer Figures \ref{fig:results1core}(c) and \ref{fig:results2core}(c)). Thus, most L2 cache lines see  less than two consecutive reads and hence, keeping three copies does not generally avoid an extra restore operation compared to keeping two copies. Further, SHIELD3 may have higher complexity than SHIELD due to keeping two and three copies of data of different widths.

\subsubsection{Effect of cache size}
To evaluate the impact of cache size, we experiment with half and double of default L2 cache size (4MB for single-core and 8MB for dual-core).

In general, restore operations incur larger overhead at lower cache sizes due to increased cache contention and hence, the scope of energy/performance gains with lower-sized caches is also higher. Hence, SHIELD brings larger gains at smaller cache sizes (refer Table \ref{tab:sensitivityresults}). Overall,  SHIELD performs consistently well at different cache sizes:  maintaining performance close to baseline, saving energy,  avoiding nearly half of the restores and bringing large reduction in write traffic.

 \section{Conclusion and future work}\label{sec:conclusion}
 In this paper, we presented a technique to mitigate read-disturbance errors in STT-RAM caches. Our technique uses compression and selective duplication of compressed-data to address both RDE and write overhead issue in STT-RAM. Experimental results performed with single and dual-core system configurations have shown that our technique is effective in improving performance and energy efficiency in presence of RDE. Also, it outperforms two other techniques for mitigating RDE. 
 
 We now list some possible directions for future work: 
 \begin{enumerate}
 \item 
 Our future work will focus on synergistic integration of SHIELD with device-level techniques (e.g., \cite{fong2014failure}) to reduce RDEs and approximate computing techniques to tolerate RDEs \cite{mittal2016surveyApproximate}.
 \item By reducing the read accesses to an STT-RAM cache, the read disturbance errors can be reduced. For achieving this, cache bypassing \cite{} or read buffers can be used. Also, using 
 
 \item We also plan to explore SRAM-STTRAM hybrid caches (e.g., \cite{mittal2015ayushCAL,agarwal2016restricting}) where frequently-accessed blocks are migrated to SRAM. This will reduce pressure on STT-RAM ways and mitigate the impact of RDE on performance. 
 
 \item In this paper, we explored STT-RAM for designing CPU caches.  CPUs primarily focus on optimizing latency in serial applications, and hence, they use large caches and small register file. By comparison, GPUs focus on optimizing throughput and hence, they use small caches and a very large register file to support their massively multithreaded architecture. Due to its high density, STT-RAM is suitable for designing GPU register file, however, the read-disturbance error   may make this  challenging. Some recent techniques seek to address this issue \cite{zhang2016red}. We plan to propose novel techniques for managing RDE in STT-RAM based GPU register file for improving performance and energy efficiency.

 \item Several other emerging memory technologies provide attractive properties. For example,  SOT-RAM (spin-orbit torque RAM) which does not suffer from read disturbance issue \cite{mittal2017SOTRAM}. We also plan to explore use of domain wall memory for designing caches \cite{mittal2016surveyDWM}.

 \end{enumerate}


\ifCLASSOPTIONcaptionsoff
  \newpage
\fi
{
\large
\bibliographystyle{IEEEtran}
\bibliography{References}

\begin{thebibliography}{10}
\providecommand{\url}[1]{#1}
\csname url@samestyle\endcsname
\providecommand{\newblock}{\relax}
\providecommand{\bibinfo}[2]{#2}
\providecommand{\BIBentrySTDinterwordspacing}{\spaceskip=0pt\relax}
\providecommand{\BIBentryALTinterwordstretchfactor}{4}
\providecommand{\BIBentryALTinterwordspacing}{\spaceskip=\fontdimen2\font plus
\BIBentryALTinterwordstretchfactor\fontdimen3\font minus
  \fontdimen4\font\relax}
\providecommand{\BIBforeignlanguage}[2]{{%
\expandafter\ifx\csname l@#1\endcsname\relax
\typeout{** WARNING: IEEEtran.bst: No hyphenation pattern has been}%
\typeout{** loaded for the language `#1'. Using the pattern for}%
\typeout{** the default language instead.}%
\else
\language=\csname l@#1\endcsname
\fi
#2}}
\providecommand{\BIBdecl}{\relax}
\BIBdecl

\bibitem{vetter2015opportunities}
J.~S. Vetter and S.~Mittal, ``{Opportunities for Nonvolatile Memory Systems in
  Extreme-Scale High Performance Computing},'' \emph{Computing in Science and
  Engineering}, 2015.

\bibitem{kaushik2017next}
B.~K. Kaushik, S.~Verma, A.~A. Kulkarni, and S.~Prajapati, \emph{Next
  Generation Spin Torque Memories}.\hskip 1em plus 0.5em minus 0.4em\relax
  Springer, 2017.

\bibitem{chun2013scaling}
K.~C. Chun, H.~Zhao, J.~D. Harms, T.-H. Kim, J.-P. Wang, and C.~H. Kim, ``{A
  scaling roadmap and performance evaluation of in-plane and perpendicular MTJ
  based STT-MRAMs for high-density cache memory},'' \emph{IEEE Journal of
  Solid-State Circuits}, vol.~48, no.~2, pp. 598--610, 2013.

\bibitem{mittal2014TPDSNVM}
S.~Mittal, J.~S. Vetter, and D.~Li, ``{A Survey Of Architectural Approaches for
  Managing Embedded DRAM and Non-volatile On-chip Caches},'' \emph{IEEE
  Transactions on Parallel and Distributed Systems (TPDS)}, 2014.

\bibitem{takemura2010highly}
R.~Takemura, T.~Kawahara, K.~Ono, K.~Miura, H.~Matsuoka, and H.~Ohno,
  ``{Highly-scalable disruptive reading scheme for Gb-scale SPRAM and
  beyond},'' in \emph{IEEE International Memory Workshop (IMW)}, 2010, pp.
  1--2.

\bibitem{mittal2017surveySoftErrorNVM}
S.~Mittal, ``A survey of soft-error mitigation techniques for non-volatile
  memories,'' \emph{Computers}, vol.~6, no.~8, 2017.

\bibitem{kang2014readability}
W.~Kang, Y.~Cheng, Y.~Zhang, D.~Ravelosona, and W.~Zhao, ``{Readability
  challenges in deeply scaled STT-MRAM},'' in \emph{Non-Volatile Memory
  Technology Symposium (NVMTS)}, 2014, pp. 1--4.

\bibitem{zhang2012prospect}
Y.~Zhang, W.~Wen, and Y.~Chen, ``{The prospect of STT-RAM scaling from
  readability perspective},'' \emph{IEEE Transactions on Magnetics}, vol.~48,
  no.~11, pp. 3035--3038, 2012.

\bibitem{ran2015read}
Y.~Ran, W.~Kang, Y.~Zhang, J.-O. Klein, and W.~Zhao, ``{Read disturbance issue
  for nanoscale STT-MRAM},'' in \emph{IEEE Non-Volatile Memory System and
  Applications Symposium (NVMSA)}, 2015, pp. 1--6.

\bibitem{wang2015selective}
R.~Wang, L.~Jiang, Y.~Zhang, L.~Wang, and J.~Yang, ``{Selective restore: an
  energy efficient read disturbance mitigation scheme for future STT-MRAM},''
  in \emph{Design Automation Conference}, 2015, p.~21.

\bibitem{jiang2016improving}
L.~Jiang, W.~Wen, D.~Wang, and L.~Duan, ``{Improving Read Performance of
  STT-MRAM based Main Memories through Smash Read and Flexible Read},''
  \emph{ASP-DAC}, 2016.

\bibitem{mittal2015compressionSurvey}
S.~Mittal and J.~Vetter, ``{A Survey Of Architectural Approaches for Data
  Compression in Cache and Main Memory Systems},'' \emph{IEEE Transactions on
  Parallel and Distributed Systems (TPDS)}, 2015.

\bibitem{chakraborty2010mc2}
A.~Chakraborty, H.~Homayoun, A.~Khajeh, N.~Dutt, A.~Eltawil, and F.~Kurdahi,
  ``{E $<$ MC2: less energy through multi-copy cache},'' in \emph{International
  conference on Compilers, architectures and synthesis for embedded systems},
  2010, pp. 237--246.

\bibitem{pekhimenko2012base}
G.~Pekhimenko, V.~Seshadri, O.~Mutlu, P.~B. Gibbons, M.~A. Kozuch, and T.~C.
  Mowry, ``Base-delta-immediate compression: practical data compression for
  on-chip caches,'' in \emph{PACT}, 2012, pp. 377--388.

\bibitem{fong2014failure}
X.~Fong, Y.~Kim, S.~H. Choday, and K.~Roy, ``{Failure mitigation techniques for
  1T-1MTJ spin-transfer torque MRAM bit-cells},'' \emph{IEEE Transactions on
  Very Large Scale Integration (VLSI) Systems}, vol.~22, no.~2, pp. 384--395,
  2014.

\bibitem{mittal2015nvmflashsurvey}
S.~Mittal and J.~Vetter, ``{A Survey of Software Techniques for Using
  Non-Volatile Memories for Storage and Main Memory Systems},'' \emph{IEEE
  Transactions on Parallel and Distributed Systems (TPDS)}, vol.~27, no.~5, pp.
  1537--1550, 2016.

\bibitem{agarwal2017towards}
S.~Agarwal and H.~K. Kapoor, ``Towards a better lifetime for non-volatile
  caches in chip multiprocessors,'' in \emph{International Conference on VLSI
  Design (VLSID)}.\hskip 1em plus 0.5em minus 0.4em\relax IEEE, 2017, pp.
  29--34.

\bibitem{wang2017decongest}
R.~Wang, S.~Mittal, Y.~Zhang, and J.~Yang, ``{Decongest: Accelerating
  Super-Dense PCM under Write Disturbance by Hot Page Remapping},'' \emph{IEEE
  Computer Architecture Letters (CAL)}, 2017.

\bibitem{mittal2015reliabilitysurvey}
S.~Mittal and J.~Vetter, ``{A Survey of Techniques for Modeling and Improving
  Reliability of Computing Systems},'' \emph{IEEE Transactions on Parallel and
  Distributed Systems}, 2015.

\bibitem{mittal2016reliabilitytradeoffs}
S.~Mittal and J.~Vetter, ``Reliability tradeoffs in design of volatile and
  non-volatile caches,'' \emph{Journal of Circuits, Systems, and Computers},
  2016.

\bibitem{seyedzadeh2016leveraging}
S.~M. Seyedzadeh, R.~Maddah, A.~Jones, and R.~Melhem, ``{Leveraging ECC to
  Mitigate Read Disturbance, False Reads and Write Faults in STT-RAM},'' in
  \emph{International Conference on Dependable Systems and Networks
  (DSN)}.\hskip 1em plus 0.5em minus 0.4em\relax IEEE, 2016, pp. 215--226.

\bibitem{kang2013high}
W.~Kang, W.~Zhao, J.-O. Klein, Y.~Zhang, C.~Chappert, and D.~Ravelosona, ``High
  reliability sensing circuit for deep submicron spin transfer torque magnetic
  random access memory,'' \emph{Electronics Letters}, vol.~49, no.~20, pp.
  1283--1285, 2013.

\bibitem{kong2016novel}
J.~Kong, ``{A novel technique for technology-scalable STT-RAM based L1
  instruction cache},'' \emph{IEICE Electronics Express}, vol.~13, no.~11, pp.
  20\,160\,220--20\,160\,220, 2016.

\bibitem{mittal2017addressing}
S.~Mittal, J.~Vetter, and L.~Jiang, ``{Addressing Read-disturbance Issue in
  STT-RAM by Data Compression and Selective Duplication},'' \emph{IEEE Computer
  Architecture Letters}, 2017.

\bibitem{zhang2016red}
H.~Zhang, X.~Chen, N.~Xiao, F.~Liu, and Z.~Chen, ``{Red-Shield: Shielding Read
  Disturbance for STT-RAM Based Register Files on GPUs},'' in \emph{Great Lakes
  Symposium on VLSI}, 2016, pp. 389--392.

\bibitem{kang2014variation}
W.~Kang, Z.~Li, J.-O. Klein, Y.~Chen, Y.~Zhang, D.~Ravelosona, C.~Chappert, and
  W.~Zhao, ``Variation-tolerant and disturbance-free sensing circuit for deep
  nanometer stt-mram,'' \emph{IEEE Transactions on Nanotechnology}, vol.~13,
  no.~6, pp. 1088--1092, 2014.

\bibitem{bishnoi2014read}
R.~Bishnoi, M.~Ebrahimi, F.~Oboril, and M.~B. Tahoori, ``{Read disturb fault
  detection in STT-MRAM},'' in \emph{IEEE International Test Conference (ITC)},
  2014, pp. 1--7.

\bibitem{raychowdhury2013pulsed}
A.~Raychowdhury, ``{Pulsed READ in spin transfer torque (STT) memory bitcell
  for lower READ disturb},'' in \emph{International Symposium on Nanoscale
  Architectures (NANOARCH)}, 2013, pp. 34--35.

\bibitem{na2016read}
T.~Na, J.~P. Kim, S.~H. Kang, and S.-O. Jung, ``Read disturbance reduction
  technique for offset-canceling dual-stage sensing circuits in deep
  submicrometer stt-ram,'' \emph{IEEE Transactions on Circuits and Systems II:
  Express Briefs}, vol.~63, no.~6, pp. 578--582, 2016.

\bibitem{mittal2016softErrorGLSVLSI}
S.~Mittal and J.~Vetter, ``Reducing soft-error vulnerability of caches using
  data compression,'' in \emph{ACM Great Lakes Symposium on VLSI (GLSVLSI)},
  Boston, Massachusetts, USA, 2016.

\bibitem{mittal2014surveycache}
S.~Mittal, ``A survey of architectural techniques for improving cache power
  efficiency,'' \emph{Elsevier Sustainable Computing: Informatics and Systems},
  vol.~4, no.~1, pp. 33--43, 2014.

\bibitem{wu2006analysis}
X.~Wu, F.~Wang, and Y.~Xie, ``{Analysis of subthreshold FinFET circuits for
  ultra-low power design},'' in \emph{IEEE International SOC Conference}, 2006,
  pp. 91--92.

\bibitem{mittal2015equalwrites}
S.~Mittal and J.~S. Vetter, ``{EqualWrites: Reducing Intra-set Write Variations
  for Enhancing Lifetime of Non-volatile Caches },'' \emph{IEEE Transactions on
  VLSI Systems}, vol.~24, no.~1, pp. 103--114, 2016.

\bibitem{poremba2015destiny}
M.~Poremba, S.~Mittal, D.~Li, J.~S. Vetter, and Y.~Xie, ``{DESTINY: A Tool for
  Modeling Emerging 3D NVM and eDRAM caches},'' in \emph{Design Automation and
  Test in Europe}, 2015.

\bibitem{mittal2014destiny}
S.~Mittal, M.~Poremba, J.~Vetter, and Y.~Xie, ``{Exploring Design Space of 3D
  NVM and eDRAM Caches Using DESTINY Tool},'' Oak Ridge National Laboratory,
  USA, Tech. Rep. ORNL/TM-2014/636, 2014.

\bibitem{mittal2016technique}
S.~Mittal and J.~Vetter, ``{A Technique For Improving Lifetime of Non-volatile
  Caches using Write-minimization},'' \emph{Journal of Low Power Electronics
  and Applications}, vol.~6, no.~1, 2016.

\bibitem{mittal2016processvariation}
S.~Mittal, ``{A Survey Of Architectural Techniques for Managing Process
  Variation},'' \emph{ACM Computing Surveys}, vol.~48, no.~4, pp. 54:1--54:29,
  2016.

\bibitem{mittal2016surveyApproximate}
S.~Mittal, ``A survey of techniques for approximate computing,'' \emph{ACM
  Computing Surveys}, vol.~48, no.~4, pp. 62:1--62:33, 2016.

\bibitem{mittal2015ayushCAL}
S.~Mittal and J.~Vetter, ``{AYUSH: A Technique for Extending Lifetime of
  SRAM-NVM Hybrid Caches},'' \emph{Computer Architecture Letters}, 2015.

\bibitem{agarwal2016restricting}
S.~Agarwal and H.~K. Kapoor, ``Restricting writes for energy-efficient hybrid
  cache in multi-core architectures,'' in \emph{IFIP/IEEE International
  Conference on Very Large Scale Integration (VLSI-SoC)}, 2016, pp. 1--6.

\bibitem{mittal2017SOTRAM}
S.~Mittal, R.~Bishnoi, F.~Oboril, H.~Wang, M.~Tahoori, A.~Jog, and J.~Vetter,
  ``{Architecting SOT-RAM Based GPU Register File},'' in \emph{IEEE Computer
  Society Annual Symposium on VLSI (ISVLSI)}, 2017.

\bibitem{mittal2016surveyDWM}
S.~Mittal, ``A survey of techniques for architecting processor components using
  domain wall memory,'' \emph{ACM Journal on Emerging Technologies in Computing
  Systems}, vol.~13, no.~2, p.~29, 2016.

\end{thebibliography}
}

\end{document}